# Sensor Fusion-based GNSS Spoofing Attack Detection Framework for Autonomous Vehicles

Sagar Dasgupta[1], Mizanur Rahman[1], Mhafuzul Islam[2], Mashrur Chowdhury[2]*

[1] Department of Civil, Construction and Environmental Engineering, University of Alabama, Tuscaloosa, AL35487, USA
[2] Glenn Department of Civil Engineering, Clemson University, Clemson, SC 29634, USA
*Senior Member, IEEE

**Abstract—**In this study, a sensor fusion based GNSS spoofing attack detection framework is presented that consists of three concurrent strategies for an autonomous vehicle (AV): (i) prediction of location shift, (ii) detection of turns (left or right), and (iii) recognition of motion state (including standstill state). Data from multiple low-cost in-vehicle sensors (i.e., accelerometer, steering angle sensor, speed sensor, and GNSS) are fused and fed into a recurrent neural network model, which is a long short-term memory (LSTM) network for predicting the location shift, i.e., the distance that an AV travels between two consecutive timestamps. We have then combined k-Nearest Neighbors (k-NN) and Dynamic Time Warping (DTW) algorithms to detect turns using data from the steering angle sensor. In addition, data from an AV's speed sensor is used to recognize the AV's motion state including the standstill state. To prove the efficacy of the sensor fusion-based attack detection framework, attack datasets are created for three unique and sophisticated spoofing attacks—turn-by-turn, overshoot, and stop—using the publicly available real-world Honda Research Institute Driving Dataset (HDD). Our analysis reveals that the sensor fusion-based detection framework successfully detects all three types of spoofing attacks within the required computational latency threshold.

**Index Terms—** Global Navigation Satellite System (GNSS), Autonomous vehicle, Cybersecurity, Spoofing attack, LSTM.

## I. INTRODUCTION

Autonomous vehicles (AVs) require accurate and reliable localization data from a Global Navigation Satellite System (GNSS) or Global Positioning System (GPS) to perform their autonomous and navigational functions effectively. The GNSS mostly depends on satellites and radio communications, which are subject to various unintentional interferences, such as buildings and thick clouds, and intentional threats, such as radio interference, communication of GNSS receiver channels, and disruptions to the GNSS infrastructures [1]-[2]. Among the foregoing, spoofing is the most sophisticated type of attack, as an attacker can mimic the authentic GNSS signal and transmit inaccurate location information to an AV [2]. Moreover, an authentic GNSS signal can be destroyed by creating a high-powered spoofing signal. For example, a spoofing signal having a power 4 dB higher than an authentic signal can destroy the authentic signal [3].

One of the primary purposes of manipulating the GNSS during a spoofing attack is to generate a spoofed signal so that an AV can be misguided to a wrong destination. For example, during a turn-by-turn type of spoofing attack, a significant shift of location could direct an AV to the wrong route. Moreover, under other types of attack, an AV can perceive that it is taking a turn or stopped while in reality, it is not doing it. Although an AV can use multiple GNSS receiver antennas to continuously cross-validate the signal, during a sophisticated spoofing attack, these antennas can be locked by an attacker into the spoofed signal using multiple phase-locked spoofers. As identified from the literature, all the current GNSS spoofing attack detection and mitigation methods are based on GNSS signal-level, and they are not adequate for detecting all types of spoofing attacks [1]-[3] . Thus, it is a challenge to develop a unified detection model against sophisticated spoofing attacks.

As the navigation of AVs for roadway transportation systems is constrained by existing roadway networks, the use of low-cost in-vehicle sensors (e.g., accelerometer, steering angle sensor, and speed sensor) could provide effective strategies for detecting sophisticated spoofing attacks. In this study, we have developed a framework for detecting spoofing attacks on GNSS services by identifying and predicting AV states, which include location shift, turning direction, and motion state using data from low-cost in-vehicle sensors. We have created attack datasets and evaluated the efficacy of our detection framework using the Honda Research Institute Driving Dataset (HDD) [4] for three sophisticated spoofing attacks: turn-by-turn, overshoot, and stop.

The rest of the paper is arranged as follows. Section II introduces the information regarding the dataset used to create the attack dataset and the data preparation approach. GNSS spoofing attack models are presented in Section III. Section IV presents our GNSS spoofing attack detection framework, analyses, and corresponding results.

## II. DATA PREPARATION

The HDD [4] is used in this study to develop and evaluate the GNSS attack detection framework. The HDD contains data from the camera, LiDAR, GPS, inertial measurement unit (IMU), and controller area network (CAN) of a conventional vehicle, and it is collected from suburban and urban roadways as well as highways within the San Francisco Bay Area. As AVs are equipped with cameras, IMU, GPS, the HDD is suitable for developing technologies for AVs. Figure 1 shows a sample route from the HDD. In HDD, the GNSS signals are recorded at 120Hz using a GeneSys Eletronik GmbH Automotive Dynamic Motion Analyzer with a Differential Global Positioning System (DGPS). The acceleration (m$^2$/s), steering wheel angle (deg), rotational speed of the steering wheel (deg/s), vehicle speed (ft/s), brake pressure (kPa), and yaw rate (deg/s) are collected from different sensors and recorded from the vehicle CAN bus at 100Hz. For our detection framework development and evaluation, the latitude and longitude, relative accelerator pedal position (%) (which represents the acceleration of an AV), steering wheel angle, and speed data are extracted from the HDD. We have then synchronized the extracted HDD by keeping GPS UNIX timestamps as references by interpolating between two closest observations. We have also calculated the perceived location shift i.e., the distance traveled between two consecutive timestamps with data from GNSS using the Haversine formula [5]:

$$d = 2r \sin^{-1}(\sqrt{\sin^2\left(\frac{\varphi_2 - \varphi_1}{2}\right) + \cos(\varphi_1)\cos(\varphi_2)\sin^2(\frac{\psi_2 - \psi_1}{2})}) \quad (1)$$

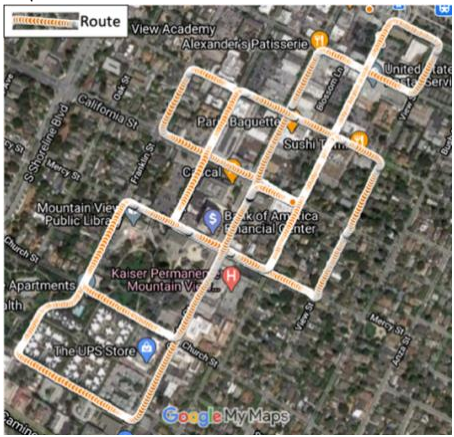

Fig.1. An example of GNSS traces from the HDD.

where *d* is the distance in meter between two points on the Earth's surface; *r* is the Earth's radius (6378 km); $\varphi_1$ and $\varphi_2$ are the latitudes in radians; $\psi_1$ and $\psi_2$ are the longitudes in radians of two consecutive time stamps.

## III. SPOOFING ATTACK MODELS

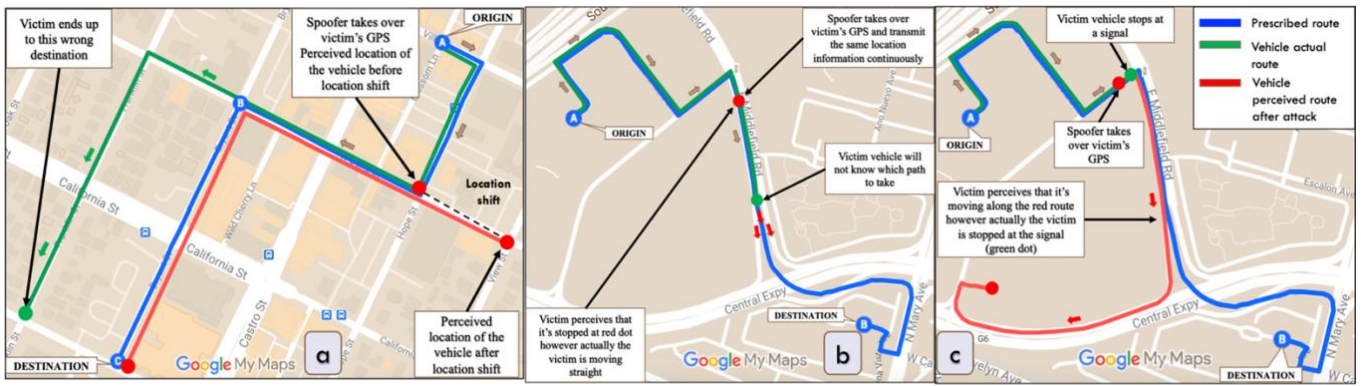

Fig.2. GNSS spoofing attack models: (a) example of turn-by-turn; (b) example of overshoot; and (c) example of stop.

During a turn-by-turn type spoofing attack [6], the spoofer transmits a synthetic signal, and the GNSS receiver locks onto the spoofed signal. After the GNSS is locked on the spoofer's signal, the spoofer changes the receiver's perceived location, resulting in a location shift between the location before and after the attack. Figure 2(a) illustrates a turn-by-turn attack in which the correct route from the origin to destination is shown in blue, the AV's ground truth route is shown in green, and the AV's perceived route, which matches with the original route turn-by-turn, is shown in red. Thus, compromising the GNSS, a spoofer creates a wrong route matching for the new route's number of turns and guides the vehicle to a wrong destination by compromising the AV's GNSS.

Figure 2(b) shows an overshoot attack [7]. After taking over the GNSS receiver, the spoofer keeps sending the same location signal to the receiver. As a result, based on the GNSS, the AV perceives that it is in a standstill state (stopped at the red dot), but the AV is moving forward in reality. When a road split occurs at the green dot, the AV will be unable to identify the path to proceed. A stop attack [7] (Figure 2(c)) is the opposite of an overshoot attack. The spoofer takes over the receiver GNSS when the AV is stopped at a stop sign (green dot) and then transmits a synthetic signal so that the AV perceives that it is moving along the road (red route). We have created ten attack scenarios for each of these three unique attack types using data from the HDD to mimic the actual spoofing scenarios at a location level.

## IV. DEVELOPMENT OF GNSS SPOOFING ATTACK DETECTION FRAMEWORK

We have developed a GNSS spoofing attack detection framework (see Figure 3), and this framework involves three concurrent strategies in which data from in-vehicle low-cost sensors—i.e., GNSS, accelerometer, steering angle, and speed—are fused to provide a unified spoofing attack detection strategy. The three strategies include the development of the attack detection framework are: (i) prediction of an AV's state information (location shift); (ii) detection of turning direction (right and left); and (iii) detection of an AV's motion state.

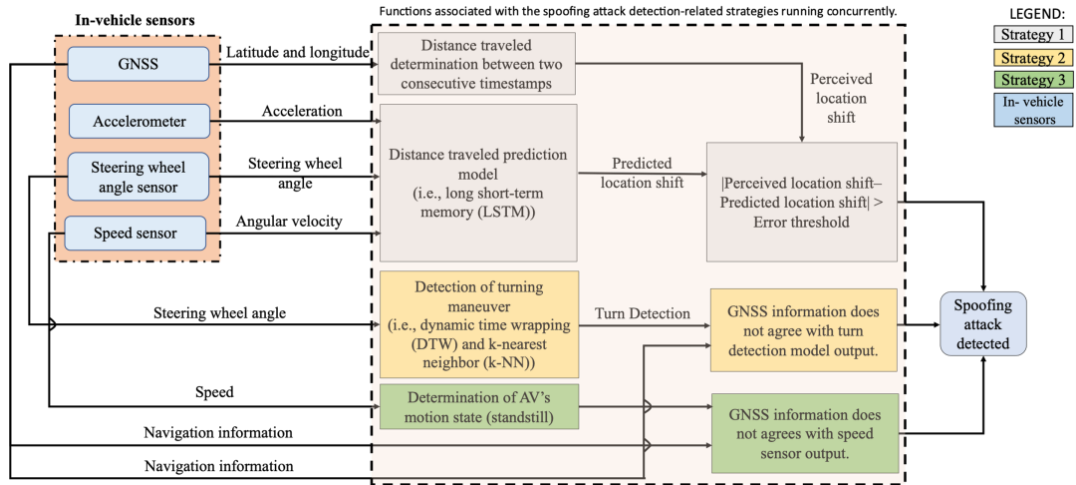

Fig. 3. Sensor fusion based GNSS spoofing attack detection framework.

The first strategy incorporates a 2-stacked long short-term memory model (LSTM) [8] with 128 and 64 neurons in the first and second hidden layers, respectively, which predicts location shift by the AV between two consecutive timestamps. The training and validation data includes the location shift between two consecutive timestamps, acceleration, steering angle, and speed. The output is the location shift between the current timestamp and the immediate future timestamp (0.00833s). Before feeding the sensor output to the LSTM training, the input features are normalized between 0 and 1. The dataset is split into training (241,990 observations) and validation (103,709 observations) datasets. The LSTM hyperparameters, i.e., number of neurons, number of epochs, batch size, and learning rate, are selected by a trial-and-error approach [8] as it is a time series prediction model. The Mean Absolute Error (MAE) metric is used as the loss function for evaluating overfitting and underfitting while training the model. The values of hyperparameters are listed in Table 1. After testing the prediction model, we found that the Root Mean Square Error (RMSE) for the predicted location shift is 0.0278 m, and the maximum absolute error is 0.0446 m. A threshold is established by adding the maximum error and GNSS positioning error as given in equation (2). Based on the GNSS device model used for collecting the HDD, the max positioning error is 0.1 m [4]. Therefore, the error threshold value is 0.1446 m and obtained by adding the maximum absolute error of the prediction model and maximum GNSS positioning error. If the difference between the perceived location shift using GNSS output between two consecutive timestamps and predicted location shift is greater than the error threshold, an attack will be detected.

$$Error\ Threshold = Prediction\ Model\ Maximum\ Absolute\ Error + Positioning\ Error\ of\ the\ GNSS \quad (2)$$

In the second strategy (turn detection), the vehicle turning maneuver is detected by comparing the steering angle sensor output and the GNSS output. If a turn is detected using steering angle data, but GNSS shows no turn, then an attack will be detected. Moreover, if the steering angle change represents a right turn and GNSS detects a left turn, an attack will be detected. The AV steering angle readings create unique shapes for left and right turns (as shown in the top left corner of Figure 4). For example, for taking a right turn, an AV first turns the steering wheel right to get into the road and then turns the steering wheel to the left to straighten the vehicle creating a unique shaped curve (see Figure 4). However, due to varying steering behavior and road geometry, the time taken for a turn is not uniform.

Table 1. LSTM model hyperparameters.

| Hyperparameters and Optimizer | Value |
|---|---|
| Number of neurons (1st layer) | 128 |
| Number of neurons (2nd layer) | 64 |
| Number of epochs | 500 |
| Batch size | 50 |
| Learning rate | 0.01 |
| Optimizer | Adam |

A k-Nearest Neighbors (k-NN) clustering algorithm is combined with a dynamic time warping (DTW) algorithm for developing left and right turn detection model using steering wheel angle data. The DTW compares the patterns and measures the similarity between two different time-series data of the different number of observations [10]. The DTW iteratively warps the time axis to align two input time series and searches for an optimal match; it then calculates the warp path distance, which is the cumulative distance between each pair of observations. The path with minimum total cost represents the DTW distance between two different time series data as shown in (3):

$$DTW(T,S) = \underset{w = w_1, w_2 \ldots, w_k, \ldots, w_K}{argmin} \sqrt{\sum_{k=1, w_k=(i,j)}^{K}(t_i - s_j)^2} \quad (3)$$

where $T$ and $S$ are ground truth and training steering angle data respectively; $w$ represents a warping path; $t_i$ is the ith observation of time series $T$; and $s_j$ is the jth observation of the time series $S$. We have used FastDTW [10] algorithm to reduce the computational time.

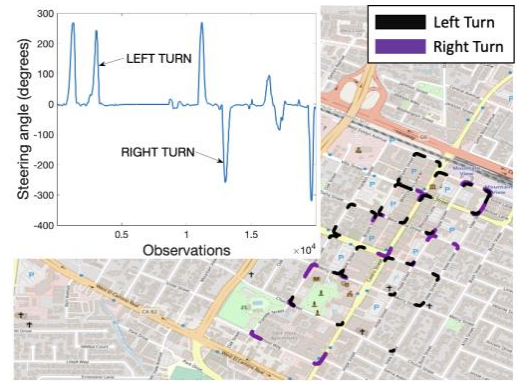

Fig. 4. Patterns of right and left turning maneuvers.

Table 2. Data used for k-NN—DTW model training

| Turn type | Dataset | Number of turns | Number of observations |
|---|---|---|---|
| Right | training | 19 | 15969 |

|      | testing  | 7  | 8209  |
|------|----------|----|-------|
| Left | training | 13 | 12974 |
|      | testing  | 6  | 6706  |

*k*-NN is a widely used classifier. The k-NN algorithm assigns a common class among its *k* nearest neighbors based on a distance metric. In this study, DTW is used as the distance metric for k-NN. The steering angle reading of right and left turns from the HDD is used for training and testing the k-NN model. The training dataset contains 19 right turns and 13 left turns, and the testing dataset contains 7 right turns and 6 left turns (see Table 2). The sample left and right turn locations using black and purple lines are shown in Figure 4. The combined *k*-NN—DTW turn detection model can classify left and right turns with an accuracy of 100% along with precision value, recall value, and F1-score of 1. Note that our training dataset contains different patterns of left and right turns.

The vehicle motion state detection strategy is useful for a spoofing attack scenario if an AV is at a standstill state and the GNSS shows the vehicle is moving. However, if a spoofer can manipulate the GNSS signal so that the perceived location shift in each timestamp is within the error threshold value, the first strategy will not be able to detect an AV's standstill state. Using the third strategy, we can detect an AV's standstill state by comparing the speed sensor's data with the GNSS speed data.

Figure 5 presents profiles for the absolute difference between the perceived location shift and predicted location shift for ten different turn-by-turn type spoofing attack scenarios (AS). At the point of beginning of the spoofing attack (as shown using different markers), the difference between perceived and predicted location shift is higher than the error threshold value; thus, it detects the attack. Note that no false attack is detected using our detection framework (see Figure 5).

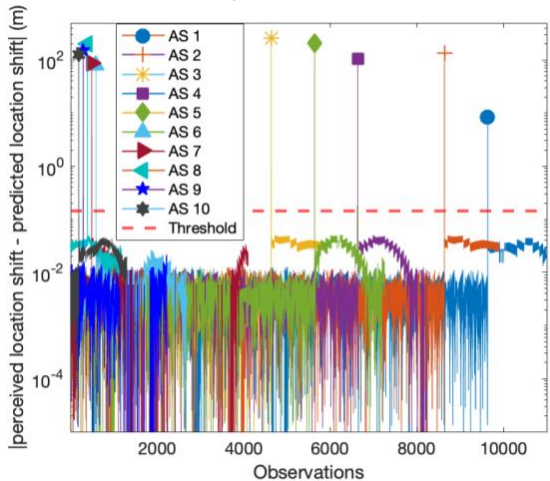

Fig. 5. Attack detection for ten turn-by-turn attack scenarios (AS).

Table 3 presents the attack evaluation matrix for all three types of spoofing attacks considered in this study. The turn-by-turn attack is detected using the location shift prediction strategy. Both the location shift prediction and vehicle motion state detection strategies are simultaneously used to detect the overshoot attack. Suppose the spoofer manages to update the location within the error threshold. Even in this case, the attack is detected due to the combination of strategies presented here can detect a standstill state (third strategy). So, all three detection strategies are required to detect the stop attack.

Table 3. Attack detection evaluation matrix

| Attack Techniques | Detection Strategies | | |
|---|---|---|---|
| | Location Shift Prediction | Vehicle Motion State Detection | Turn Detection |
| Turn-by-turn | ✓ | NA | NA |
| Overshoot | ✓ | ✓ | NA |
| Stop | ✓ | ✓ | ✓ |
| ✓- Detected, NA - Not applicable | | | |

The average computational latency for our first strategy—i.e., location shift prediction strategy— is 0.691μs for each observation, which is less than the GNSS data generation frequency (i.e., 120Hz or 0.0083s or 8300μs). In our second strategy, we have resampled the steering angle sensor data to 5Hz from 120Hz because our experiments showed that 5Hz sampling rate reduces the computational time while preserving the steering patterns of right and left turns. The k-NN- DTW model takes 0.08s on average to detect a turn, which is less than the data sampling frequency (i.e., 5Hz or 0.2s). The third strategy takes 0.00034s on average to compare and detect the vehicle's motion state. Note that, we used a workstation equipped with dual Intel Xeon Gold 5215 2.5GHz processor with 128GB DDR4 2666MHz RDIMM ECC RAM memories for calculating all the above-mentioned computational time. In conclusion, our attack detection framework is able to detect different types of attacks with a high degree of success. Further research can be performed to evaluate and validate the framework through real-world experiments.

REFERENCES

[1] J. Zidan, E. I. Adegoke, E. Kampert, S. A. Birrell, C. R. Ford, and M. D. Higgins, "GNSS Vulnerabilities and Existing Solutions: A Review of the Literature," *IEEE Access*, pp. 1–1, Feb. 2020, doi: 10.1109/access.2020.2973759.
[2] S. Dasgupta, M. Rahman, M. Islam, and M. Chowdhury, "Prediction-Based GNSS Spoofing Attack Detection for Autonomous Vehicles," *Transportation Research Board*, Oct. 2020, Accessed: Feb. 25, 2021. [Online]. Available: http://arxiv.org/abs/2010.11722.
[3] Z. Wu, Y. Zhang, Y. Yang, C. Liang, and R. Liu, "Spoofing and Anti-Spoofing Technologies of Global Navigation Satellite System: A Survey," *IEEE Access*, vol. 8, pp. 165444–165496, Sep. 2020, doi: 10.1109/access.2020.3022294.
[4] Ramanishka, Y.-T. Chen, T. Misu, and K. Saenko, "Toward Driving Scene Understanding: A Dataset for Learning Driver Behavior and Causal Reasoning," *Proceedings of the IEEE Com. Society Conf. on Computer Vision and Pattern Recognition*, pp. 7699–7707, Nov. 2018.
[5] C. C. Robusto, "The Cosine-Haversine Formula," *The American Mathematical Monthly*, DOI: 10.2307/2309088.
[6] K. Zeng, Y. Shu, S. Liu, Y. Dou, and Y. Yang, "A practical GPS location spoofing attack in road navigation scenario," in *HotMobile 2017 - Proceedings of the 18th Intl. Workshop on Mobile Computing Systems and Applications*, Feb. 2017, pp. 85–90, doi: 10.1145/3032970.3032983.
[7] J. R. Van Der Merwe, X. Zubizarreta, I. Lukčin, A. Rügamer, and W. Felber, "Classification of Spoofing Attack Types," in *2018 European Navigation Conf., ENC 2018*, Aug. 2018, DOI: 10.1109/EURONAV.2018.8433227.
[8] Z. Khan, M. Chowdhury, M. Islam, C. Y. Huang, and M. Rahman, "Long Short-Term Memory Neural Network-Based Attack Detection Model for In-Vehicle Network Security," *IEEE Sensors Letters*, vol. 4, no. 6, Jun. 2020, doi: 10.1109/LSENS.2020.2993522.
[9] S. Salvador and P. Chan, "FastDTW: Toward Accurate Dynamic Time Warping in Linear Time and Space." Accessed: Apr. 22, 2021. [Online].